\documentclass[aps,prl,twocolumn,showpacs,groupedaddress]{revtex4}
\usepackage{graphicx}
\usepackage{epstopdf}
\begin{document}

\title{Local structure of supercritical matter}
\author{Dima Bolmatov$^{1}$}
\thanks{d.bolmatov@gmail.com, db663@cornell.edu}
\author{D. Zav'yalov$^{2}$}
\author{M. Zhernenkov$^{3}$}

\address{$^1$ Baker Laboratory, Cornell University, Ithaca, New York 14853-1301, USA}
\address{$^2$ Volgograd State Technical University, Volgograd, 400005 Russia}
\address{$^3$ Brookhaven National Laboratory, Upton, NY 11973, USA}

\begin{abstract}
The supercritical state is currently viewed as uniform on the pressure-temperature phase diagram. Supercritical fluids have the dynamic motions of a gas but are able to dissolve materials like a liquid. They have started to be deployed in many important industrial applications stimulating fundamental theoretical work and development of experimental techniques. Here, we have studied local structure of supercritical matter by calculating static structure factor, mean force potential, self-diffusion, first coordination shell number and pair distribution function within very wide temperature ranges. Our results show a monotonic disappearance of medium-range order correlations at elevated temperatures providing direct evidence for structural crossover in the reciprocal and real spaces. Importantly, the discovered structural crossover in the reciprocal space is fundamentally inter-related to structural crossover in the real space, granting new unexpected interlinks between operating system properties in the supercritical state. Finally, we discuss an evolution analysis of the local structure and important implications for an experimental detection of structural monotonic transitions in the supercritical matter.
\end{abstract}

\maketitle
In studies of materials, periodic arrangement of the atoms plays very crucial role \cite{ashcroft1}. This periodicity, and its underlying symmetry, has effects on the formal description of properties such as vibrational frequencies and energy spectra \cite{kittel1,bol1}. Liquids and supercritical fluids are distinguished from solids by their lack of long-range structural order \cite{ziman1,bol2,bol3}. Studies of fluids can only focus on the arrangements of neighbouring groups of atoms which define local structure. Indeed, analysis at this level often yields surprising insights into the structure of soft matter and even crystalline materials. Thus local structure is important for a wide range of condensed matter and materials science with the ongoing effort in understanding the local order and thermodynamic properties of disordered matter such as supercritical fluids \cite{trachenko1}, classical \cite{hopkins1,dyre1,hangjun1,jani1,bolmatov2,anisimov1,tan2} and quantum liquids \cite{boljap}.

One of the principal tools in the theoretical study of soft matter is the method of molecular dynamics simulations (MD). This computational method calculates the time dependent behavior of a system. MD simulations provide detailed information on the fluctuations and conformational changes of atoms and molecules. MD is now routinely used to investigate the structure, dynamics and thermodynamics of fluids and other complex materials. The varying degrees of disorder in complex materials also requires experimental techniques, which go beyond the routine crystallographic structure solutions of Bragg scattering which only reveals the average long-range structure. The use of diffuse scattering to determine the local atomic arrangements gives a key to understanding the behavior of modern complex materials as well as possibly predicting their properties. Current modern techniques are based on the analysis of total scattering patterns, which includes both Bragg and diffuse scattering, and provides comprehensive information on the local atomic structure of various materials. In recent years, there has been impressive progress in developing and applying novel experimental techniques \cite{,egami1,loerting1,dolan1,oganov1,loerting2,prakap1}.
\begin{figure}
	\centering
\includegraphics[scale=0.47]{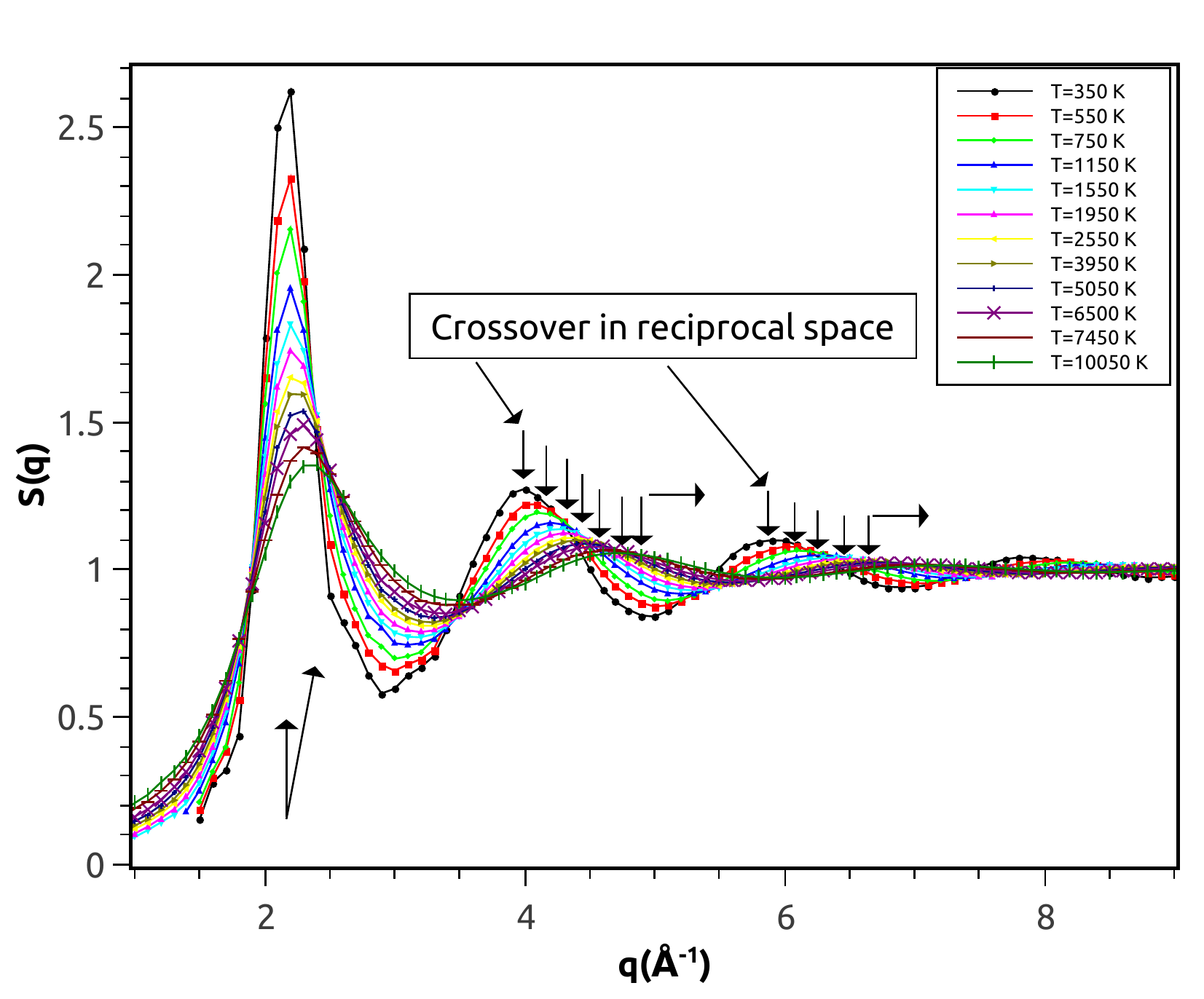}
\caption{Structural evolution in reciprocal space. Evolution of structure factor $S(q)$ of simulated one-component Lennard-Jones (LJ) supercritical fluid at different temperatures showing the disappearance of the medium-range order at high temperature. The simulations are performed at constant density.}
	\label{fig1}
\end{figure}

\begin{figure*}[htp]
  \centering
  \begin{tabular}{cc}
    
    \includegraphics[width=61mm]{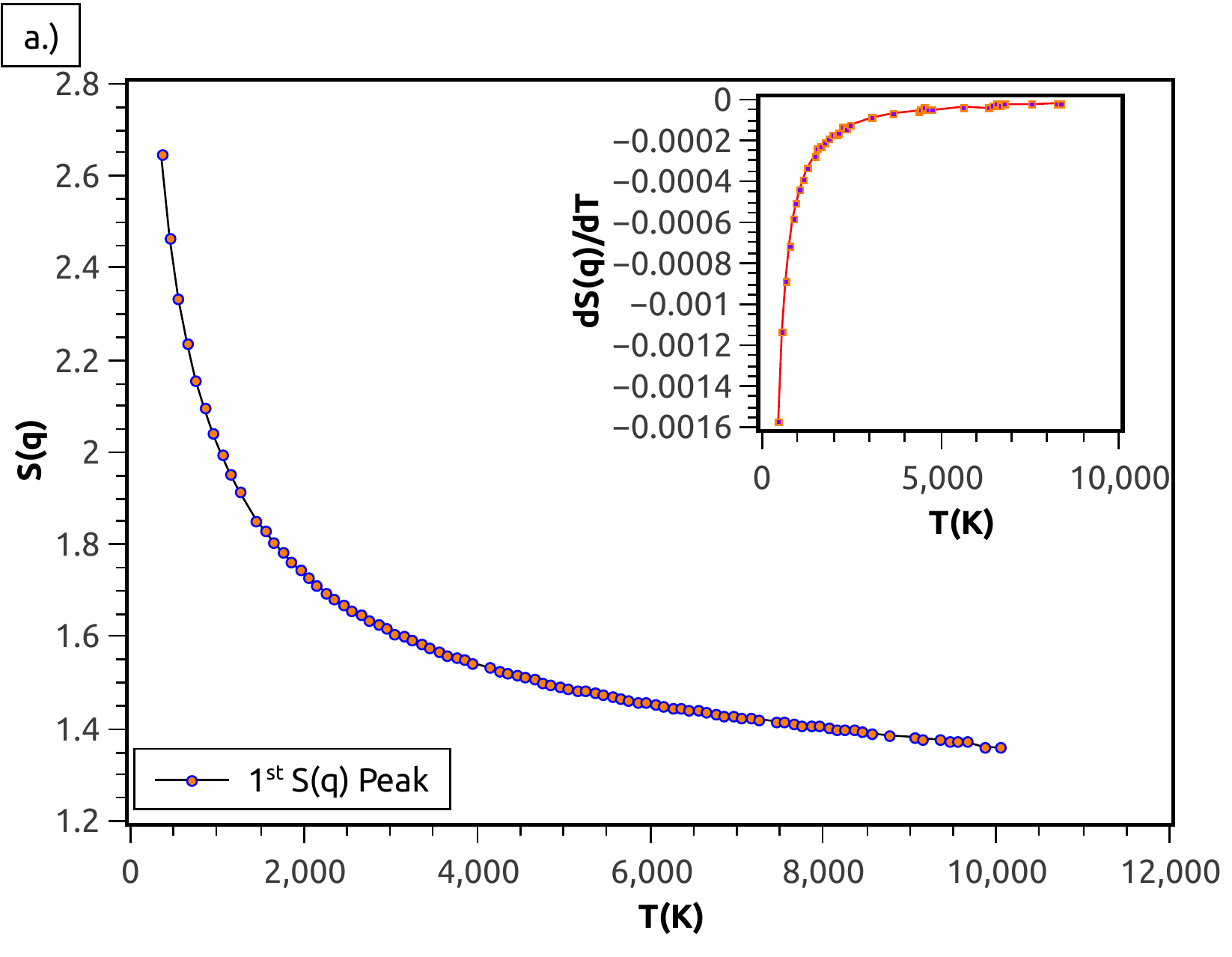}&

    \includegraphics[width=61mm]{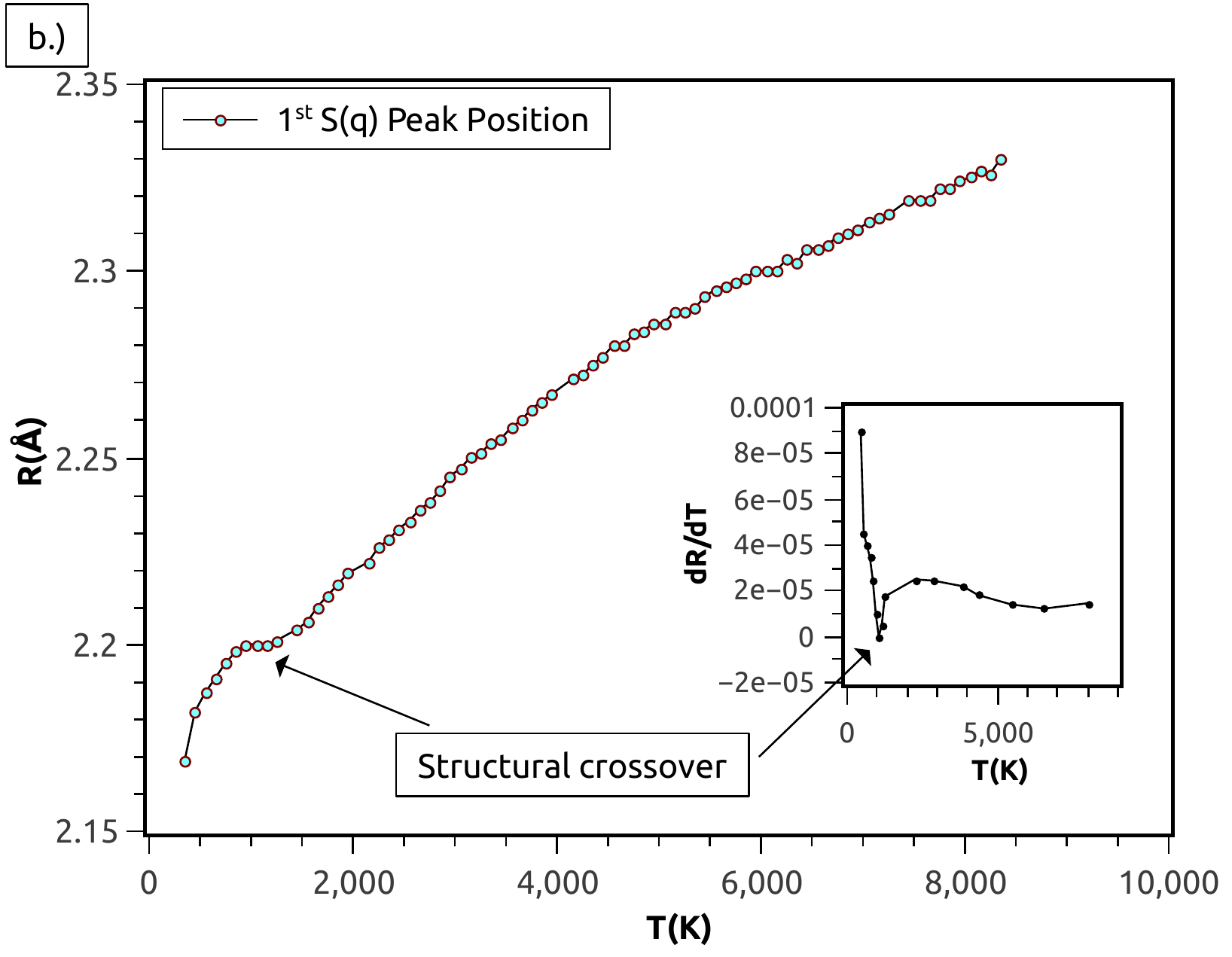}\\

    \includegraphics[width=61mm]{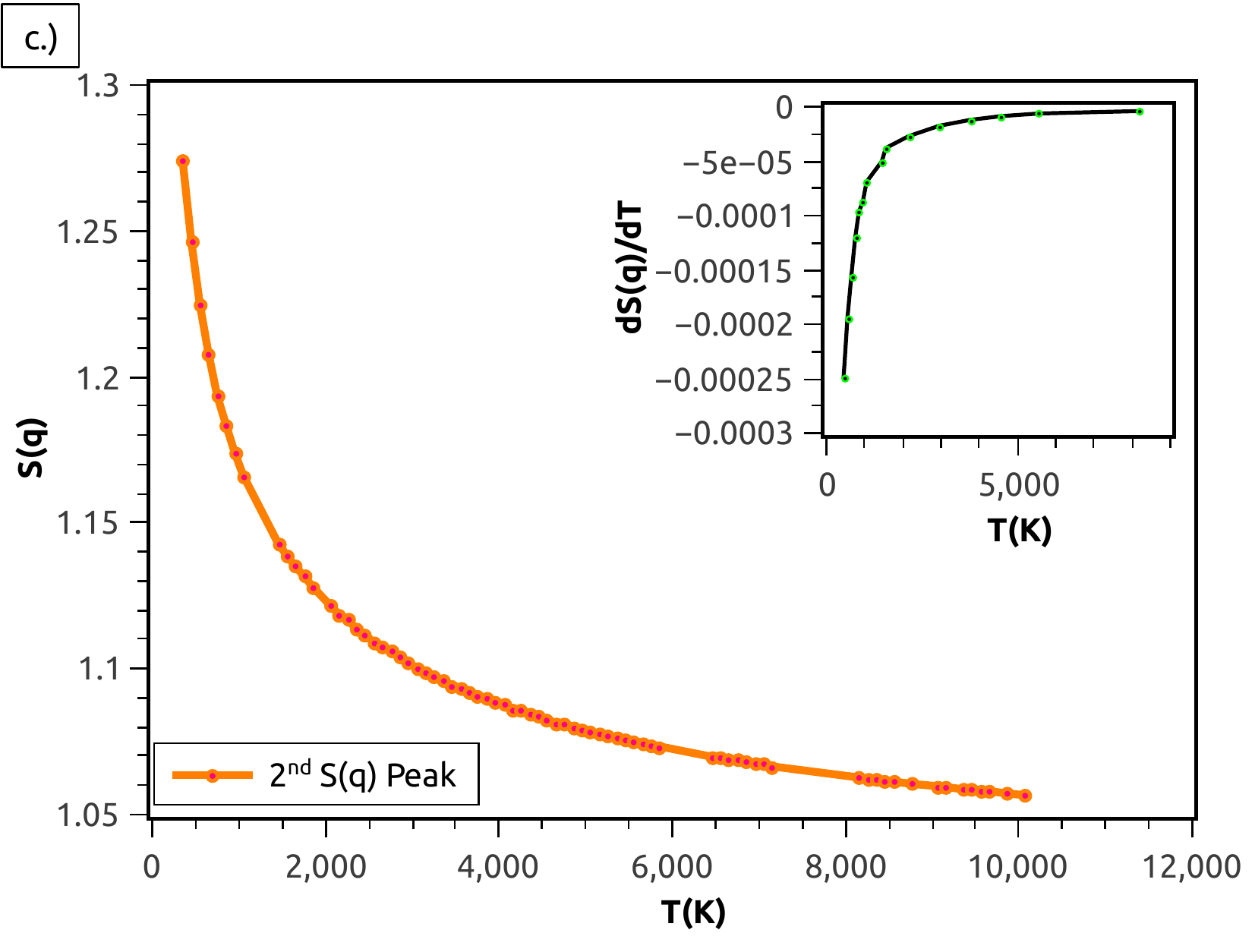}&

    \includegraphics[width=61mm]{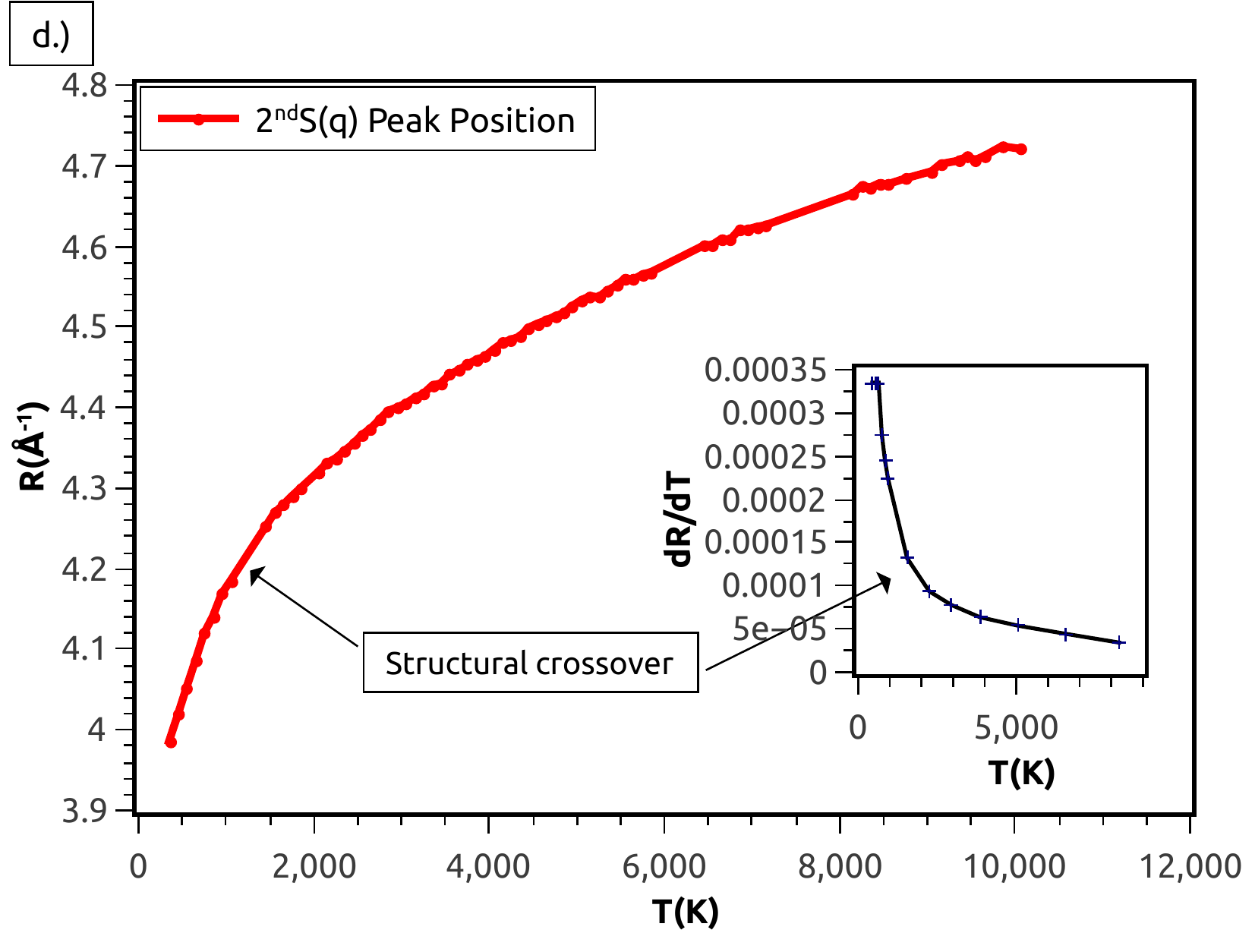}\\

  \end{tabular}
 \caption{Structural evolution in reciprocal and real spaces. Evolution of first and second $S(q)$ peaks and their positions at different temperatures. Insets show $S(q)$ peaks and their positions derivatives.}
 \label{figur}
\end{figure*}

Using MD simulations, we have simulated one-component Lennard-Jones (LJ) fluid fitted to Ar properties \cite{martin}. We have simulated the system with 32000 atoms using constant-volume (NVE) ensemble in the very wide temperature range (see Figs. 1--3) well extending into the supercritical region; the system was equilibrated at constant temperature. The temperature range in Figs. 1-3 is between about 3$T_c$ and 167$T_c$, where $T_c$  is the critical temperature of Ar, $T_{c}\simeq$150 K (1.3 in LJ units). The simulated density, 1880 kg/m$^3$ (1.05 in LJ units), corresponds to approximately three times the critical density of Ar. A typical MD simulation lasted about 50 ps, and the properties were averaged over the last 20 ps of simulation, preceded by 30 ps of equilibration. The simulations at different temperature included 100 temperature points simulated on the high-throughput computing cluster. We calculate pair distribution function $g(r)$ (PDF), static structure factor $S(q)$ and diffusion, average it over the last 20 ps of the simulation, and show its temperature evolution in Figures 1--3. We observe the decrease of the first peak of $S(q)$, and the near disappearance of the second and third peaks, implying that the medium-range order correlations is no longer visible at high temperature.

In order to explore the local structure of supercritical state, we calculate static structure factor $S(q)$ from corresponding PDF. The  structure factor $S(q)$ can be defined as:
\begin{eqnarray}
S(q) = 1 + 4\pi\varrho \int_{{0}}^{{R_{max}}}dr r^2{\frac{{\sin{qr}}}{{qr}}} \left({g(r)-1}\right)
\end{eqnarray}
where the $g(r)$ is the pair distribution function \cite{ashcroft1} and $R_{max}=$ 20 $\AA$ we used in our calculations. 

In the limiting case of no interaction, the system is an ideal gas and the structure factor is completely featureless: $S(q)=1$. Because there is no correlation between the positions $\mathbf{r}_j$ and $\mathbf{r}_k$ of different particles (they are independent random variables), so the off-diagonal terms in equation: 
\begin{eqnarray}
S(q) = 1 + \frac{1}{N} \left \langle \sum_{j \neq k} \mathrm{e}^{-i \mathbf{q} (\mathbf{r}_j - \mathbf{r}_k)} \right \rangle
\end{eqnarray}
average to zero
\begin{eqnarray}
\langle \exp [-i \mathbf{q} (\mathbf{r}_j - \mathbf{r}_k)]\rangle = \langle \exp (-i \mathbf{q} \mathbf{r}_j) \rangle \langle \exp (i \mathbf{q} \mathbf{r}_k) \rangle = 0
\end{eqnarray}
Even for interacting particles, at high scattering vector the structure factor goes to 1. This result follows from equation: $S(q) = 1 + \rho \int_V \mathrm{d} \mathbf{r} \, \mathrm{e}^{-i \mathbf{q}\mathbf{r}} g(r)$,
since $S(q)-1$ is the Fourier transform of the "regular" function $g(r)$ and thus goes to zero for high values of the argument $q$. This reasoning does not hold for a perfect crystal, where the distribution function exhibits infinitely sharp peaks.
\begin{figure*}[htp]
  \centering
  \begin{tabular}{cc}
    
    \includegraphics[width=61mm]{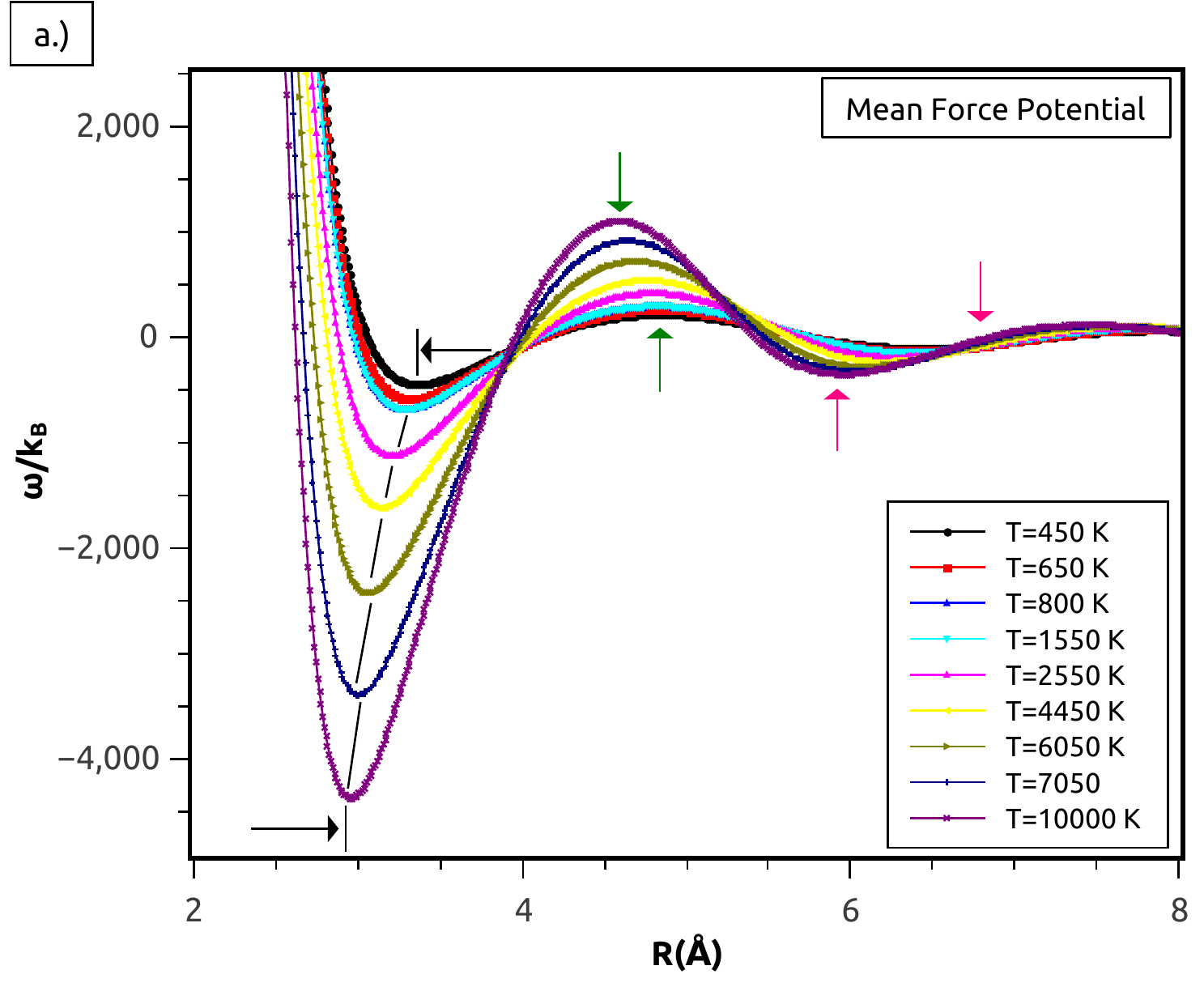}&

    \includegraphics[width=61mm]{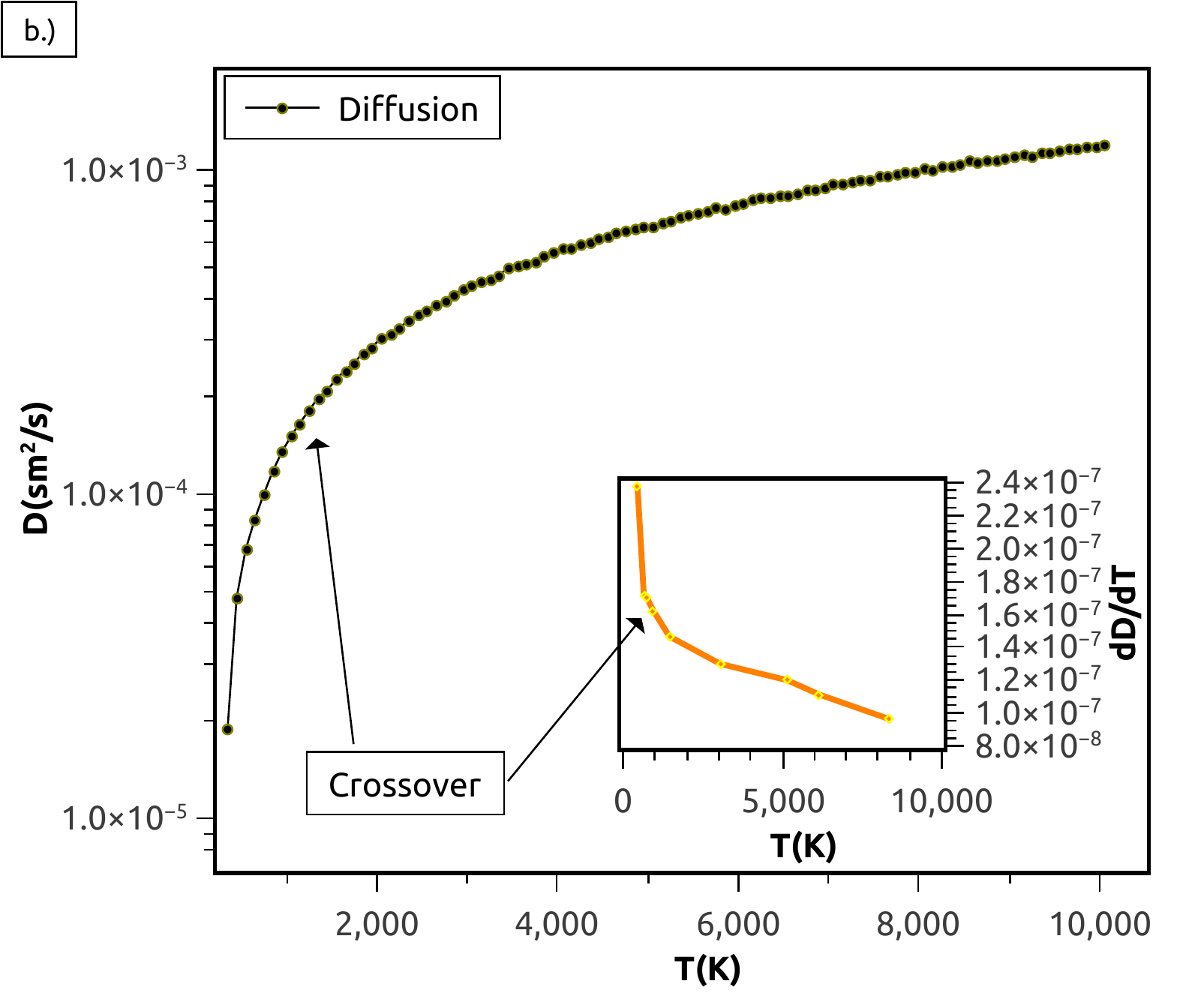}\\

    \includegraphics[width=61mm]{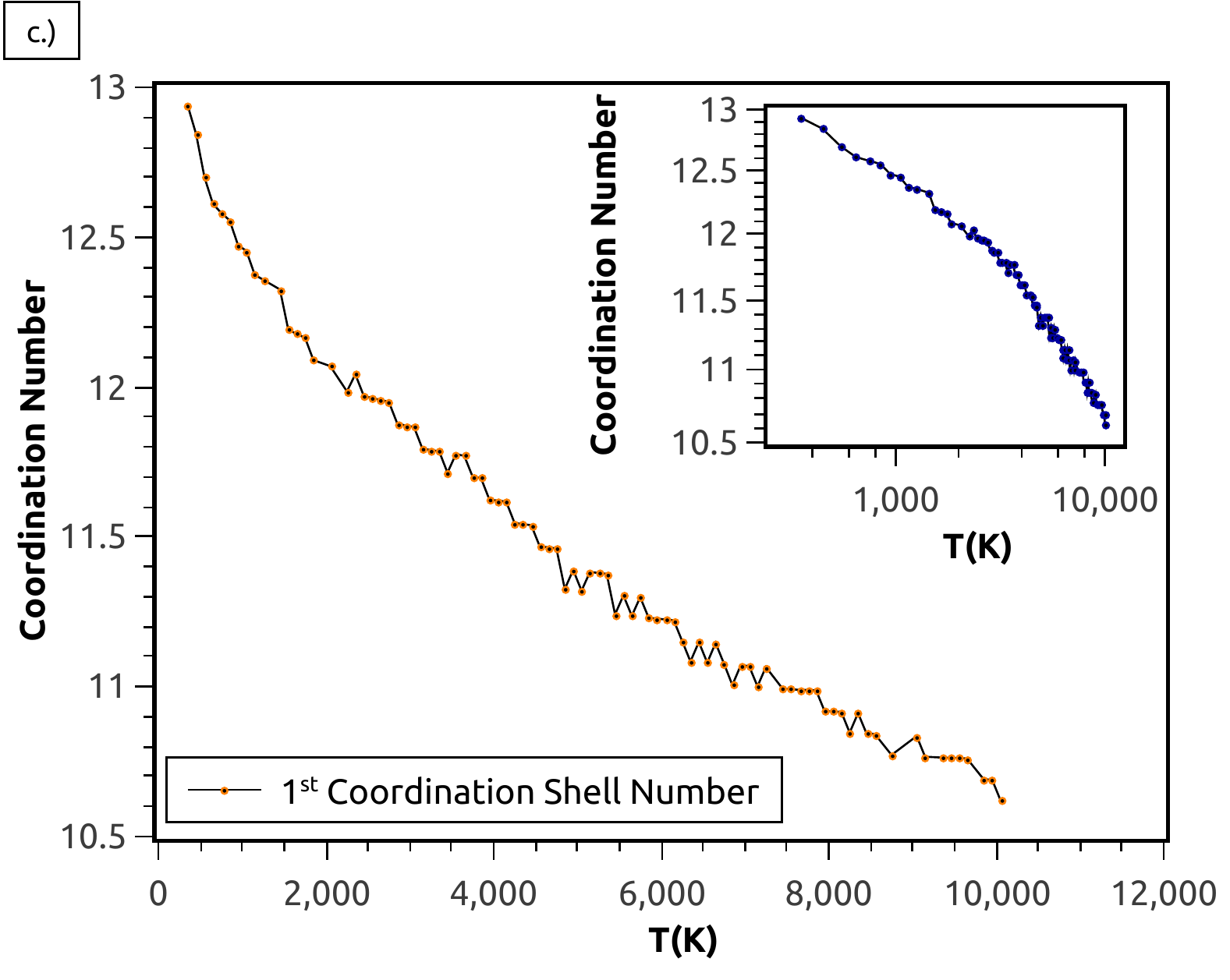}&

    \includegraphics[width=61mm]{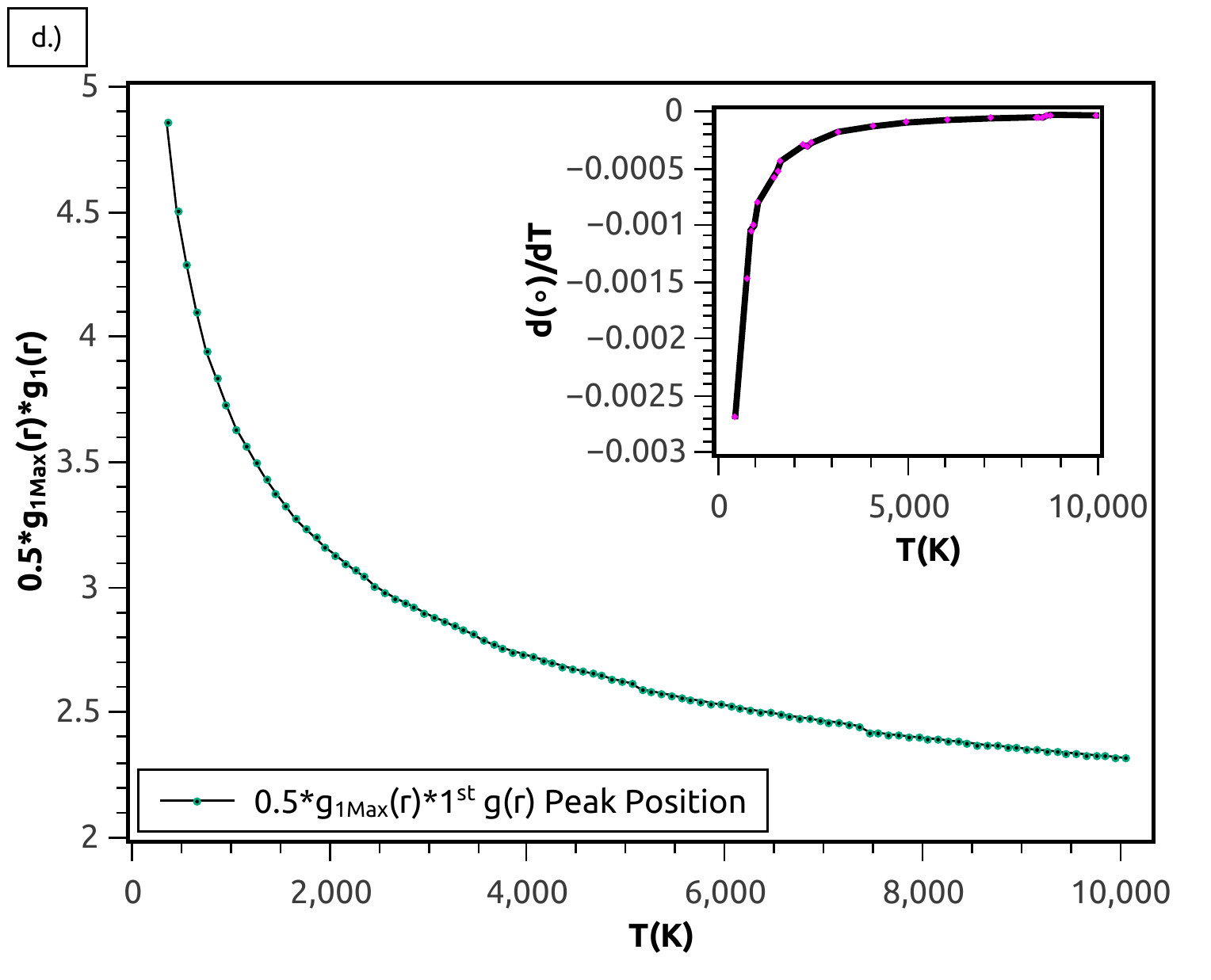}\\

  \end{tabular}
 \caption{Evolution of mean force potential, diffusion and first coordination shell at different temperatures. Insets show structure factor peak and their position derivatives. Fig.3b inset shows the derivative of self-diffusion with respect to temperature. Fig. 3c inset shows the coordination number in logarithmic scale. In Fig. 3d $(\circ)=0.5\times g_{1Max}(r)\times$ $1^{st}$ peak position.}
 \label{figur}
\end{figure*}

To analyze temperature changes of $S(q)$ in more detail (see Fig. 1), we calculate the heights and positions of the first and second peaks of $S(q)$, and plot these in Figure 2 as a function of temperature. We observe the steep decrease of both peaks at low temperature (Fig. 2a and Fig. 2c), followed by their flattening at high temperature, with the crossover between the two regimes taking place around 1000 K. To make the crossover more visible, in inserts (see Fig. 2) we plot the temperature derivative of the heights and positions of both peaks. These plots clearly show two regimes corresponding to the fast and slow change of $S(q)$ peaks and their positions. The structural crossover in the supercritical state, now evidenced in the reciprocal space, is the new effect not hitherto anticipated, in view of the currently perceived physical uniformity of the supercritical state.

The structural crossover is further evidenced by the calculation of mean force potential (MFP), diffusion and first coordination shell characteristic parameters: first shell coordination number, height and their position at different temperatures (see Fig. 3). Figure 3 (evolution of system properties in the real space) shows the same trend as the Figure 2 (evolution of system properties in the real space). 

Various methods have been proposed for calculating MFP. The simplest representation of the MFP is to use the separation $r$ between two particles as the reaction coordinate. The MFP (see Fig. 3a) is related to the PDF using the following expression for the Helmholtz free energy \cite{kirkwood}: $w=-k_{\rm B}T\ln{\left[g(r)\right]}$, where $k_{\rm B}$ is the Boltzmann constant. To calculate diffusion we use a standard method, Einstein's method, which relates the mean square distance (MSD) travelled by a certain particle over a certain time interval. At the limit of observation time goes to infinity, self-diffusion in terms of MSD \cite{heyes1}: $D=\lim_{t\rightarrow\infty}\frac{1}{6Nt}\left<\sum_{i=1}^{N}[\vec{r}_{i}(t)-\vec{r}_{i}(0)]^2\right>$, where $\vec{r}_{i}$ is displacement vector of the $i$-th atom at $t$ time and the term $[\vec{r}_{i}(t)-\vec{r}_{i}(0)]^2$ is the MSD.

Although the structural order of a fluid is usually enhanced by isothermal compression or isochoric cooling,
a few notable systems show the opposite behaviours. Specifically, increasing density can disrupt the structure of liquid-like fluids, while lowering temperature or strengthening of attractive interactions can weaken the correlations of fluids with short-range attractions. It is a particularly insightful quantity to study because its contributions from the various coordination shells of the pair distribution function can be readily determined, and it correlates strongly with static structure factor behaviour (see Figs. 1--3), which allows it to provide insights into structural crossover operating in the supercritical state.

We note that the relationship between system properties in the real and inverse spaces has always fascinated scientists in condensed matter physics, with the recognition that such a relationship may exist in some classes of systems but not in others. Here, we find that not only the supercritical state is physically non-uniform as previously viewed, it is also universally amenable to supporting fundamental interlinks between system behaviour both in the reciprocal and real spaces consistently.

We now address the origin of structural crossover in the reciprocal space, and relate this origin to the changes of dynamics, thermodynamics and structure of the supercritical state. It has been confirmed before \cite{pilgrim1,sette1,inui1}, when liquid relaxation time $\tau$ (the average time between two consecutive atomic jumps at one point in space \cite{frenkel}) approaches its minimal value $\tau_{\rm D}$ , the Debye vibration period, the system loses the ability to support propagating high-frequency shear modes with $\omega >\frac{2\pi}{\tau}$ and behaves like a gas \cite{bolsr,bolprb}. When all shear modes are lost, only the longitudinal mode remains in the system, yielding heat capacity at constant volume $c_{V}=2 k_{\rm B}$ per particle \cite{bol2}. This result can be easily obtained from the phonon theory of liquids in the classical limit \cite{bolsr}. Therefore, the discovered structural crossover in the reciprocal space is also closely related to both dynamic and thermodynamic crossovers existing in the supercritical state.

An easy and elegant way to experimentally determine the structural crossover is to deduce the pair distribution function from a conventional diffraction experiment with neutrons or X-rays. The use of the structure factor in interpretation of scattering data is particularly valuable since it's directly related to pair distribution function. In diffraction experiment, energy-resolved measurements gives access to dynamic structure factor, $S(q,\omega)$.The static structure factor $S(q)$ is measured without resolving the energy of X-rays or neutrons and can be written as: $S(q) = 1 + \rho \int_V \mathrm{d} \mathbf{r} \, \mathrm{e}^{-i \mathbf{q}\mathbf{r}} g(r)$, where $\rho$ is a number density and $g(r)$ is a pair distribution function. Since the structural crossover of supercritical Argon is observed at high temperature (up to 1500 K) and high pressure (up to 5 GPa), the use of small sample volumes and sophisticated sample environment ($e.g.$ diamond anvil cells) is inevitable. Such experiments can only be performed at brightest synchrotron sources or most intense neutron beams. Thus practical significance of this investigation relies on the expectation that structural crossover both in the reciprocal and real spaces can be experimentally observed. The existence of this crossover has not been hitherto anticipated, and is contrary to how the supercritical state has been viewed until now. We believe that further theoretical and experimental investigation of fundamental interlinks between structure and dynamics in the supercritical matter can also lead to greater understanding in another condensed matter systems. 
\section{Acknowledgements} Dima Bolmatov thanks Ben Widom and Cornell University for support. We are grateful to Neil Ashcroft for fruitful discussions.

\end{document}